\documentclass[12pt,preprint]{aastex} 
\shorttitle{Orbit of Regulus} 
\shortauthors{Gies et al.} 
\begin{document} 
 
\title{A Spectroscopic Orbit for Regulus} 
 
\author{D. R. Gies\altaffilmark{1,2}, 
S. Dieterich\altaffilmark{1},   
N. D. Richardson\altaffilmark{1},   
A. R. Riedel\altaffilmark{1},  
B. L. Team\altaffilmark{1}, \\  
H. A. McAlister\altaffilmark{1},  
W. G. Bagnuolo, Jr.\altaffilmark{1},  
E. D. Grundstrom\altaffilmark{1,2,3}, \\
S. \v{S}tefl\altaffilmark{4},  
Th. Rivinius\altaffilmark{4}, and  
D. Baade\altaffilmark{5}} 
 
\altaffiltext{1}{Center for High Angular Resolution Astronomy, 
Department of Physics and Astronomy, 
Georgia State University, P. O. Box 4106, Atlanta, GA  30302-4106; 
gies@chara.gsu.edu, dieterich@chara.gsu.edu, richardson@chara.gsu.edu, 
riedel@chara.gsu.edu, team@chara.gsu.edu, hal@chara.gsu.edu, bagnuolo@chara.gsu.edu, 
erika.grundstrom@vanderbilt.edu} 
 
\altaffiltext{2}{Visiting Astronomer, Kitt Peak National Observatory, 
National Optical Astronomy Observatory, operated by the Association 
of Universities for Research in Astronomy, Inc., under contract with 
the National Science Foundation.} 
 
\altaffiltext{3}{Current address: Physics and Astronomy Department,  
Vanderbilt University, 6301 Stevenson Center, Nashville, TN 37235} 
 
\altaffiltext{4}{European Organisation for Astronomical Research  
in the Southern Hemisphere, Alonso de Cordova 3107, Vitacura, 
Santiago de Chile, Chile; sstefl@eso.org, triviniu@eso.org}  
 
\altaffiltext{5}{European Organisation for Astronomical Research  
in the Southern Hemisphere, Karl-Schwarzschild-Str. 2,  
85748 Garching bei M\"{u}nchen, Germany; dbaade@eso.org} 
 
 
 
\begin{abstract} 
We present a radial velocity study of the rapidly rotating B-star 
Regulus that indicates the star is a single-lined spectroscopic 
binary.  The orbital period (40.11~d) and probable semimajor axis 
(0.35 AU) are large enough that the system is not interacting at 
present.  However, the mass function suggests that the secondary  
has a low mass ($M_2 > 0.30 M_\odot$), and we argue that the companion 
may be a white dwarf.  Such a star would be the remnant of a former  
mass donor that was the source of the large spin angular momentum of  
Regulus itself. 
\end{abstract} 
 
 
 
\setcounter{footnote}{5} 
 
\section{Introduction}                              
 
Regulus ($\alpha$ Leo; HD~87901; HR~3982; HIP~49669) is a nearby  
($d=24.3\pm0.2$ pc; \citealt{van07}) intermediate mass star of  
spectral type B7~V \citep{joh53} or B8~IVn \citep{gra03}.   
It is one of a number of nearby B- and A-type stars exhibiting  
extremely fast rotation.  The very broad shape of its photospheric 
absorption lines indicates a projected rotational velocity of  
$V\sin i = 317 \pm 3$ km~s$^{-1}$ \citep{mca05}.  The full picture 
of its fast spin came with the first interferometric observations  
of Regulus with the CHARA Array optical long baseline interferometer 
\citep{mca05}.  These observations showed that the star is rotationally 
flattened and gravity darkened at its equator.  Models of the  
spectrum and interferometry demonstrate that the star has a rotation  
period of 15.9~hr, with an equatorial velocity equal to $86\%$ of the  
critical velocity, where centripetal acceleration balances gravity.  
 
The fast spin of Regulus is puzzling given its probable age ($\approx 150$ 
Myr; \citealt*{ger01}).  Stars born as fast rotators are expected to slow 
down relatively quickly after birth, and only again achieve  
rapid rotation at the conclusion of core H-burning through a redistribution 
of angular momentum \citep{eks08}.  Thus, it is surprising to find
rapid rotation in Regulus, a star which is still  
in the middle of its core H-burning stage.  On the other hand, stars  
that are members of interacting binaries can experience large changes  
in spin due to tidal interactions and mass exchange.  \citet{lan08}  
discuss how mass transfer may lead to the spin up of the mass gainer in  
a large fraction of these binaries.  Depending on the initial separation  
and mass ratio, the system may merge or it  
may widen following mass ratio inversion, leaving the donor remnant in  
a large orbit that shuts down mass transfer once the donor's envelope is 
lost.  We know of several examples of such post-mass transfer binaries  
with rapid rotators, including the Be X-ray binaries with neutron star  
companions \citep{coe00} and Be binaries with He-star companions  
\citep{gie98,mai05,pet08}.  
 
Regulus does have a known wide companion ($\alpha$~Leo~B 
at a separation of $\approx 175\arcsec$, which is itself a binary  
consisting of K2~V and M4~V pair; \citealt{mca05}), but this companion  
has far too great a separation to have ever interacted directly with  
Regulus.  There are no known closer companions, but the  
last significant radial velocity investigation was made in 1912 -- 1913 
by \citet{mel23}.  The scatter in the results introduced by the broad and  
shallow appearance of the spectral lines may have discouraged other  
investigators, but this early work and others \citep*{mau92,fro26,cam28,pal68} 
suggest that any velocity variations present are relatively small.  However,  
a low mass donor remnant would probably create only a modest reflex  
motion in Regulus, so the lack of demonstrated variability is not  
unexpected.  We have made spectroscopic observations of Regulus on  
many occasions over the last few years, and here we present a  
summary of the velocities measured in these spectra.  We find that  
Regulus is in fact a low amplitude, single-lined, spectroscopic  
binary, and we discuss the possible nature of the companion.  
 
 
\section{Observations and Radial Velocities}        
 
Table~1 lists the sources and properties of the spectra of  
Regulus we used to measure radial velocity.  The spectra from 
run numbers 1 -- 8 were made by us and have moderate resolving power 
and good S/N  (usually better than 100 per pixel).  These include spectra  
obtained with the Kitt Peak National Observatory Coud\'{e} Feed Telescope 
\citep{val04}, the Czech Academy of Sciences Ondrejov Observatory 
telescope and HEROS spectrograph \citep{ste00}, and 
the Multiple-Telescope Telescope at the Georgia State University  
Hard Labor Creek Observatory \citep*{bar02}.  
We have also obtained a number of spectra from on-line archives including 
the ESO La Silla 50 cm telescope and HEROS spectrograph, 
University of Toledo Ritter Observatory echelle spectrograph \citep{mor97},
the Elodie spectrograph of the Observatoire de Haute Provence \citep{mou04},
the ESO VLT and UVES (UVES Paranal Observatory Project, 
ESO DDT Program ID 266.D-5655; \citealt{bag03}), 
La Silla 3.6 m telescope and HARPS spectrograph \citep{may03}, and 
La Silla 2.2 m telescope and FEROS spectrograph \citep{kau99,wes08}.
In many cases, the archival spectra included a series made within a 
few minutes time, and we report here the average velocity of such groups. 
Finally, we also collected a series of 12 UV high dispersion 
spectra from the archive of the {\it International Ultraviolet  
Explorer (IUE)}.  All these spectra were reduced to a rectified  
continuum format using standard routines in  
IRAF\footnote{IRAF is distributed by the 
National Optical Astronomy Observatory, which is operated by 
the Association of Universities for Research in Astronomy, Inc., 
under cooperative agreement with the National Science Foundation.}, 
and then each group was transformed onto a uniform, heliocentric 
wavelength grid in increments of $\log\lambda$.  Many of these  
observations record the red spectrum in the vicinity of H$\alpha$,  
and we usually removed the atmospheric telluric lines in this  
part of the spectrum using contemporaneous spectra of rapidly  
rotating A-type stars or using the atlas of atmospheric  
transmission\footnote{ftp://ftp.noao.edu/catalogs/atmospheric\_transmission/}  
made by L.\ Wallace, W.\ Livingston, and K.\ Hinckle (KPNO).  
The {\it IUE} spectra were similarly transformed 
to a uniform $\log \lambda$ grid \citep*{pen97}.  In prior studies of more 
distant O-stars, we have checked the wavelength calibration 
by registering the positions of interstellar lines with those in the
average spectrum.  Regulus, however, is so close that most of the
interstellar lines are too weak, and in the end we relied on the 
measurement of a single feature, \ion{O}{1} $\lambda 1302$, for 
registration, which introduces some additional scatter into our 
radial velocities from the {\it IUE} spectra.  
 
\placetable{tab1}      
 
All the radial velocities were measured using the cross-correlation  
method with errors estimated according to the scheme described  
by \citet{zuc03}.  All the optical spectra were cross-correlated 
with a synthetic model spectrum taken from the work of \citet{mar05}. 
This spectrum is based upon a Kurucz model atmosphere for solar  
abundances, interpolated to $T_{\rm eff} = 12200$~K and $\log g = 3.5$,  
which are close to average values over the visible hemisphere  
\citep{mca05}.  The model template was smoothed with a rotational  
broadening function to better match the actual spectrum, and in  
each case the model was transformed to the observed wavelength grid. 
Unfortunately, the models of \citet{mar05} do not extend to UV wavelengths, 
so for the {\it IUE} spectra we used a model UV template from the  
TLUSTY/Synspec models of \citet{lan07} for $T_{\rm eff} = 15000$~K 
(the lowest temperature in their grid) and $\log g = 3.75$.  The  
use of a different template for the UV spectra may introduce  
systematic differences from those obtained from the optical spectra,  
but these errors are probably comparable to the measurement errors 
(see \S3).  
 
Our 168 measurements are gathered in Table~2 (given in full in the  
electronic version) that lists the heliocentric 
Julian date of mid-observation, the orbital phase (\S3), the  
radial velocity and its internal error, the observed minus  
calculated residual (\S3), and the run number corresponding  
the observational journal in Table~1. 
 
\placetable{tab2}      
 
 
\section{Orbital Elements}                          

The range in the radial velocities is larger than that expected 
from measurement errors, so we searched for evidence of periodic
variations using the discrete Fourier transform and CLEAN 
method \citep*{rob87} and phase dispersion minimization \citep{ste78}. 
Both procedures identified the presence of one significant 
period at $P=40.11 \pm 0.02$~d, with a power indicating a
false alarm probability of $\sim 10^{-22}$ that the peak results 
from random errors \citep{sca82}.   This period is too long to 
be related to rotational or pulsational variations, so we 
assume that it results from orbital motion in a binary. 
We then derived the remaining orbit elements using the 
nonlinear, least squares fitting program of \citet{mor74}
by keeping the orbital period fixed at the value given above. 
Each measurement was assigned a weight proportional to the 
inverse square of the larger of the measurement error
or 1~km~s$^{-1}$ (to account for possible systematic 
errors between results from different groups of observations). 
Trials with other weighting schemes gave similar results.
Elliptical solutions made no significant improvement in the
residuals from the fit \citep{luc71}, so we adopted a
circular fit.  We present in Table~3 the standard orbital 
elements where $T_0$ is the epoch of the ascending node.  
Note that the error in $T_0$ increases to $\pm 3.9$~d
when the full range in acceptable period is considered.  
The derived systemic velocity $V_0$ is similar to the
radial velocity of $\alpha$~Leo~B of $6.56\pm 0.22$ km~s$^{-1}$
\citep{tok02}, which strengthens the case for a physical connection 
in this common proper motion pair. 

\placetable{tab3}      

The radial velocity curve and measurements are illustrated in 
Figure~1.  The {\it IUE} measurements ({\it open circles}) 
show the largest scatter around the curve, which we think 
derives from errors related to registering the wavelength 
scale with a single interstellar line (\S2).  Most of the 
residuals for the optical spectra have a size comparable 
to the measurement errors and are mainly free from 
systematic trends.  However, several of the runs that 
recorded the red spectrum around H$\alpha$ do have 
residuals that are systematically low (see run $\#3$). 
We suspect that these trends are due to subtle differences 
in data treatment, but because these specific runs cover 
only a limited part of the orbital cycle, we did not apply
any corrections for systematic differences.  
 
\begin{figure}[h] 
\begin{center} 
{\includegraphics[angle=90,height=12cm]{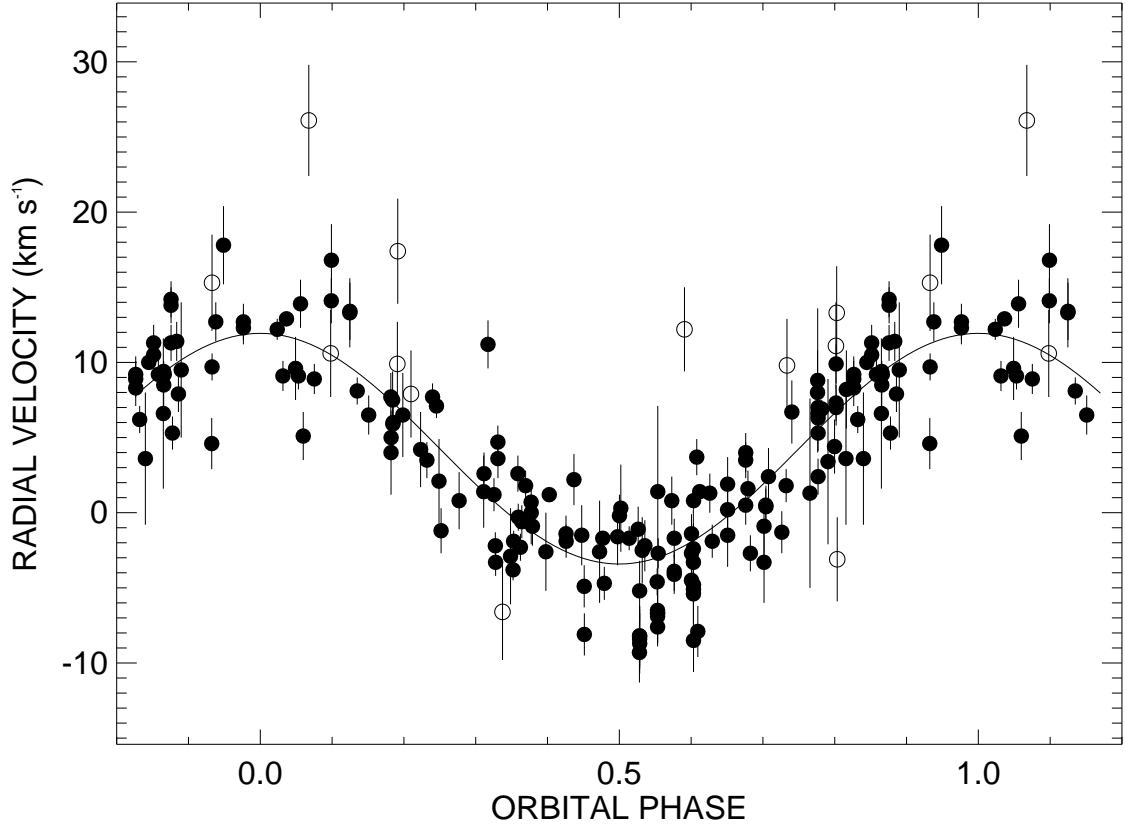}} 
\end{center} 
\caption{The observed and derived radial velocity curves.  
The open circles indicate the {\it IUE}  
UV measurements while the solid circles represent the optical  
spectra measurements.} 
\label{fig1} 
\end{figure}

\placefigure{fig1}     
 
 
\section{Discussion}                                

The orbital variation has a small semiamplitude that eluded 
detection in earlier studies.  Consequently, the derived 
mass function is also small (Table~3), and we show in Figure~2 
the constraints on the possible masses from the mass function.
\citet{mca05} used model fits to derive a probable mass of 
the primary star of $M_1 = 3.4 \pm 0.2 M_\odot$, and the 
boundaries of this range are indicated by the vertical 
dotted lines.  Larger orbital inclinations are favored for 
random orientations, and it is possible that the orbital 
inclination is comparable to the spin inclination 
of Regulus, $i\approx 90^\circ$ \citep{mca05}.  Thus, the 
mass of the companion may be close to the minimum mass 
shown (for $i= 90^\circ$) of $M_2 > 0.30\pm 0.01 M_\odot$.  
 
\begin{figure}[h] 
\begin{center} 
\plotone{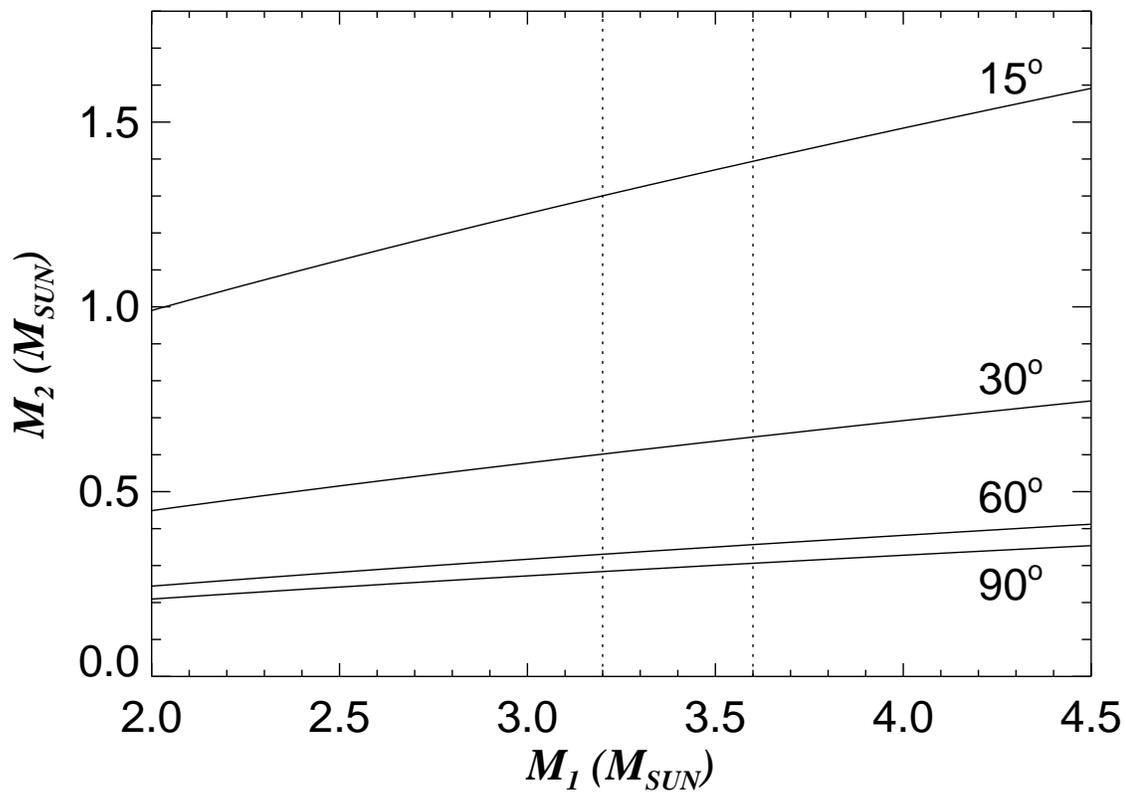}  
\end{center} 
\caption{The mass diagram constraints for the mass of  
Regulus ($M_1$) and its faint companion ($M_2$).  Each solid 
line gives the relation from the mass function for the  
orbital inclination indicated on the right hand side.  
The vertical dotted lines show the probable mass range for  
Regulus \citep{mca05}.} 
\label{fig2} 
\end{figure}

\placefigure{fig2}     

A companion this small may be a low mass white dwarf or 
main sequence star.  If Regulus was spun up by mass 
transfer in an interacting binary, then the remnant of 
the donor star is probably a low mass white dwarf \citep{rag01,wil04}.
Indeed, the lowest mass white dwarfs are usually found in 
binary systems \citep*{mar95} where they lost a significant
fraction of their mass, and some reach masses as low 
as $0.17 M_\odot$ \citep{kil07}.  Models by \citet{wil04} 
for mass transfer during H-shell burning (their ``evolutionary
channel 1'') often lead to remnant and gainer masses and 
orbital periods similar to the case of Regulus.  Since Regulus 
and its wider companion $\alpha$~Leo~B are not too old 
($<150$ Myr; \citealt{ger01}), a white dwarf companion would 
not have progressed too far along its cooling track and 
would probably have an effective temperature $>16000$~K 
\citep*{alt01}, much higher than that of the primary 
B7~V star.  Consequently, we might expect to observe 
a modest FUV flux excess if the companion is a white dwarf, 
and, in fact, \citet{mor01} find that the spectral energy 
distribution of Regulus is about a factor of two brighter
in the 1000 -- 1200~\AA ~range than predicted by model 
atmospheres for a single B7~V star.  We note for completeness
that a neutron star companion is probably ruled out 
because it would require a very small inclination (see Fig.~2)
and an unlikely evolutionary scenario in which only a modest 
amount of mass transfer occurred before the supernova explosion
(no more than the present mass of the primary). 

On the other hand, the companion could be a low mass, main 
sequence star (making Regulus one of the most extreme mass 
ratio binaries among the massive stars after exclusion of the
massive X-ray binaries).  Adopting the minimum mass, the companion
would be an M4~V star that would be too faint to alter 
significantly the spectral energy distribution from that 
for the primary alone (see Fig.~4 in \citealt{mca05}).  
It is very unlikely that the companion is the outer 
component of a once compact triple system.  
If the fast spin of Regulus is the result of 
prior mass transfer, then an M4~V star in the orbit we find 
would have been too close to the central binary for orbital 
stability (and would have likely been ejected or become a 
merger product).  Thus, in binary models for rapid rotation, 
the companion cannot be a low mass main sequence star but must 
be the remnant of the donor. 
 
The angular separation of the binary is probably small.  
If we assume masses of $M_1=3.4 M_\odot$ and $M_2=0.3 M_\odot$, 
then according to Kepler's third law, the semimajor axis is 
0.35~AU.  Thus, for a distance of 24.3~pc, the maximum 
angular separation will be approximately 15 mas, too small for
detection by speckle interferometric or adaptive optics techniques. 
The binary could be resolved in principle by lunar occultation  
methods or optical long baseline interferometry, but the fact that 
there is no evidence of a binary from these methods is consistent
with the expected faintness of the companion \citep*{han74,rad81,rid82,mca05}.  
For example, the magnitude difference in the $K$-band is probably close 
to $\triangle m \approx 10$ and 6 mag for the cases of a white dwarf 
and an M4~V star companion, respectively.   Thus, the flux of the 
companion has no influence on the analysis of the interferometry 
presented by \citet{mca05}. 

Our study has led to the discovery of a binary companion to 
the twenty second brightest star in the sky, and it may, like 
Sirius, offer another example of a bright 
star that is orbited by a faint white dwarf.   If the companion 
is a white dwarf, then it may be the closest case of a star 
stripped to its core by mass transfer in a close binary.  
The best opportunity to test the white dwarf hypothesis 
will come from very short wavelength observations where a hot
companion may outshine the B7~V primary. 
 
 
\acknowledgments 
 
We thank the staff of Kitt Peak National Observatory for  
their support in obtaining these observations.  
This work is partially based on spectral data retrieved from 
the ELODIE archive at Observatoire de Haute-Provence (OHP). 
Additional spectroscopic data were retrieved from Ritter  
Observatory's public archive, which is supported by the National  
Science Foundation Program for Research and Education with Small  
Telescopes (NSF-PREST) under grant AST-0440784.
The {\it IUE} data presented in this paper were obtained from 
the Multimission Archive at the Space Telescope Science 
Institute (MAST).  Support for MAST for non-HST data is 
provided by the NASA Office of Space Science via grant 
NAG5-7584 and by other grants and contracts.
This work was also supported by the National Science 
Foundation under grant AST-0606861 (DG). 
Institutional support has been provided from the GSU College 
of Arts and Sciences and from the Research Program Enhancement 
fund of the Board of Regents of the University System of Georgia, 
administered through the GSU Office of the Vice President 
for Research.  We are grateful for all this support. 
 

 
\clearpage 
\begin{deluxetable}{cccccl} 
\tablewidth{0pc} 
\tabletypesize{\scriptsize} 
\tablenum{1} 
\tablecaption{Journal of Spectroscopy \label{tab1}} 
\tablehead{ 
\colhead{Run} & 
\colhead{Dates} & 
\colhead{Range} & 
\colhead{Resolving Power} & 
\colhead{} & 
\colhead{Observatory/Telescope/} \\ 
\colhead{Number} & 
\colhead{(BY)} & 
\colhead{(\AA)} & 
\colhead{($\lambda/\triangle\lambda$)} & 
\colhead{$N$} & 
\colhead{Spectrograph}} 
\startdata 
 1\dotfill & 1989.3           & 4453 -- 4597               &\phn\phn6280 &\phn6 & KPNO/0.9m/Coud\'{e} \\ 
 2\dotfill & 2000.9           & 6445 -- 6700               &\phn   12100 &   30 & KPNO/0.9m/Coud\'{e} \\ 
 3\dotfill & 2004.8           & 6466 -- 6700               &\phn\phn9500 &   16 & KPNO/0.9m/Coud\'{e} \\ 
 4\dotfill & 2005.9           & 4240 -- 4580               &\phn   10300 &\phn2 & KPNO/0.9m/Coud\'{e} \\ 
 5\dotfill & 2006.8           & 6434 -- 6700               &\phn\phn7600 &\phn2 & KPNO/0.9m/Coud\'{e} \\ 
 6\dotfill & 2006.8           & 4240 -- 4580               &\phn   10200 &\phn2 & KPNO/0.9m/Coud\'{e} \\ 
 7\dotfill & 2000.9 -- 2002.4 & 3780 -- 5700, 5832 -- 8483 &\phn   20000 &   46 & Ond\v{r}ejov/2m/HEROS \\ 
 8\dotfill & 1999.1 -- 2000.3 & 6500 -- 6700               &\phn   14000 &\phn6 & HLCO/MTT 1m/Ebert-Fastie \\ 
 9\dotfill & 1999.4           & 5832 -- 8483               &\phn   20000 &\phn1 & La Silla/0.5m/HEROS \\ 
10\dotfill & 2004.2 -- 2007.1 & 6527 -- 6596               &\phn   26000 &   37 & Ritter/1m/Echelle \\ 
11\dotfill & 1996.3           & 4000 -- 5000               &\phn   34100 &\phn2 & OHP/1.9m/Elodie \\ 
12\dotfill & 2003.0           & 3760 -- 4500, 4800 -- 5100 &\phn   80000 &\phn1 & VLT/8m/UVES \\ 
13\dotfill & 2004.0           & 3800 -- 5000               &      120000 &\phn2 & La Silla/3.6m/HARPS \\ 
14\dotfill & 2006.1 -- 2007.1 & 3750 -- 5150               &\phn   48000 &\phn3 & La Silla/2.2m/FEROS \\ 
15\dotfill & 1979.0 -- 1995.3 & 1200 -- 1900               &\phn   10000 &   12 & {\it IUE}/0.45m/Echelle (SWP) \\ 
\enddata 
\end{deluxetable} 
 
\begin{deluxetable}{lccccc} 
\tabletypesize{\scriptsize} 
\tablewidth{0pt} 
\tablenum{2} 
\tablecaption{Radial Velocity Measurements \label{tab2}} 
\tablehead{ 
\colhead{Date}              & 
\colhead{Orbital}           & 
\colhead{$V_r$}             & 
\colhead{$\triangle V_r$}   & 
\colhead{$(O-C)$}           & 
\colhead{Run}               \\ 
\colhead{(HJD$-$2,400,000)} & 
\colhead{Phase}             & 
\colhead{(km s$^{-1}$)}     & 
\colhead{(km s$^{-1}$)}     & 
\colhead{(km s$^{-1}$)}     & 
\colhead{Number}            } 
\startdata 
 43881.848 \dotfill &  0.933 & 
\phs     $  15.3$ & 3.2 & \phn\phs $   4.0$ &     $ 15$ \\
 44333.437 \dotfill &  0.191 & 
\phn\phs $   9.9$ & 2.8 & \phn\phs $   2.9$ &     $ 15$ \\
 44333.458 \dotfill &  0.191 & 
\phs     $  17.4$ & 3.5 & \phs     $  10.4$ &     $ 15$ \\
 44529.050 \dotfill &  0.068 & 
\phs     $  26.1$ & 3.7 & \phs     $  14.8$ &     $ 15$ \\
 45360.745 \dotfill &  0.802 & 
\phs     $  11.1$ & 2.9 & \phn\phs $   4.4$ &     $ 15$ \\
\enddata 
\tablecomments{Table 2 is available in its entirely in the electronic edition.
A portion is shown here for guidance regarding its form and content.}
\end{deluxetable} 
 
 
\begin{deluxetable}{lc} 
\tablewidth{0pc} 
\tablenum{3} 
\tablecaption{Circular Orbital Elements\label{tab3}} 
\tablehead{ 
\colhead{Element}                         & \colhead{Value}       } 
\startdata 
$P$ (d)                          \dotfill & 40.11 (2)\tablenotemark{a} \\ 
$T_0$ (HJD -- 2,400,000)         \dotfill & 44526.3 (3)                \\ 
$K_1$ (km s$^{-1}$)              \dotfill & 7.7 (3)                    \\ 
$V_0$ (km s$^{-1}$)              \dotfill & 4.3 (2)                    \\ 
$a_1\sin i$ ($R_\odot$)          \dotfill & 6.1 (3)                    \\ 
$f(M)$ ($M_\odot$)               \dotfill & 0.0019 (2)                 \\ 
$(M_2 \sin i) / (1+M_2/M_1)^{2/3}$ ($M_\odot$) \dotfill & 0.279 (14)     \\
r.m.s. (km s$^{-1}$)             \dotfill & 2.8                        \\ 
\enddata 
\tablenotetext{a}{Fixed.} 
\tablecomments{Numbers in parentheses give the error in the last digit quoted.} 
\end{deluxetable} 
 
 
\input{epsf} 
 
\clearpage 
 
\clearpage 
 
\clearpage 
 
 
\center{\large ONLINE MATERIAL}

{\tt
\begin{verbatim}
Title: A Spectroscopic Orbit for Regulus
Authors: Gies et al. 
Table: Radial Velocity Measurements
================================================================================
Byte-by-byte Description of file: datafile2.txt
--------------------------------------------------------------------------------
   Bytes Format Units   Label  Explanations
--------------------------------------------------------------------------------
   1- 10 F10.3  d       Date   HJD-2400000
  11- 16 F6.3   --      Phase  Orbital phase from ascending node
  17- 21 F5.1   km/s    HRV    Heliocentric radial velocity
  22- 26 F5.1   km/s  e_HRV    Mean error on HRV
  27- 31 F5.1   km/s    O-C    Observed minus calculated residual
  32- 34 I3     --      Run    Run number from Table 1
--------------------------------------------------------------------------------
 43881.848 0.933 15.3  3.2  4.0 15
 44333.437 0.191  9.9  2.8  2.9 15
 44333.458 0.191 17.4  3.5 10.4 15
 44529.050 0.068 26.1  3.7 14.8 15
 45360.745 0.802 11.1  2.9  4.4 15
 45360.787 0.803 13.3  3.1  6.5 15
 45360.817 0.803 -3.1  2.8 -9.9 15
 45377.118 0.210  7.9  2.9  1.7 15
 46184.477 0.337 -6.6  3.2 -6.9 15
 46876.553 0.591 12.2  2.8 14.4 15
 47645.697 0.766  1.3  6.3 -3.7  1
 47646.702 0.791  3.4  5.5 -2.8  1
 47647.706 0.816  3.6  4.4 -3.7  1
 47648.682 0.840  3.6  4.4 -4.8  1
 47649.686 0.865  6.6  5.0 -2.7  1
 47650.685 0.890  9.5  4.5 -0.7  1
 47899.708 0.098 10.6  2.9  0.1 15
 49810.478 0.734  9.8  3.1  6.3 15
 50204.407 0.554 -2.7  1.0  0.3 11
 50206.381 0.603  0.8  0.6  2.7 11
 51197.705 0.317 11.2  1.6 10.1  8
 51254.792 0.740  6.7  2.1  2.9  8
 51307.569 0.056 13.9  1.6  2.4  9
 51603.630 0.437  2.2  1.7  5.0  8
 51650.663 0.609 -7.9  1.7 -6.2  8
 51654.610 0.708  2.4  1.9  0.2  8
 51663.620 0.932  4.6  1.7 -6.7  8
 51860.664 0.845 10.0  0.4  1.4  7
 51890.018 0.576 -3.9  1.3 -1.4  2
 51890.019 0.576 -4.1  1.3 -1.6  2
 51890.019 0.576 -1.7  1.3  0.8  2
 51890.982 0.600 -1.4  1.3  0.5  2
 51890.982 0.600 -4.5  1.3 -2.6  2
 51890.982 0.600 -2.7  1.3 -0.8  2
 51892.005 0.626  1.3  1.3  2.4  2
 51893.001 0.651  0.2  2.0  0.4  2
 51893.001 0.651  1.9  1.8  2.1  2
 51893.002 0.651 -1.5  2.1 -1.3  2
 51894.011 0.676  0.5  1.3 -0.3  2
 51894.011 0.676  3.5  1.2  2.7  2
 51894.012 0.676  4.0  1.3  3.2  2
 51896.029 0.726 -1.3  1.4 -4.4  2
 51898.054 0.777  6.6  1.3  1.1  2
 51898.054 0.777  6.3  1.2  0.8  2
 51898.054 0.777  2.4  1.2 -3.1  2
 51898.055 0.777  7.0  1.2  1.5  2
 51898.055 0.777  5.3  1.2 -0.2  2
 51899.060 0.802  7.3  1.2  0.6  2
 51899.061 0.802  7.0  1.2  0.3  2
 51899.061 0.802  9.9  1.3  3.2  2
 51899.634 0.816  8.2  2.6  0.8  7
 51900.047 0.826  8.9  1.2  1.1  2
 51900.047 0.826  9.2  1.2  1.4  2
 51900.048 0.827  8.3  1.2  0.5  2
 51901.048 0.851 11.3  1.2  2.5  2
 51901.048 0.851 10.5  1.2  1.7  2
 51901.603 0.865  9.4  0.4  0.1  7
 51901.612 0.865  8.5  1.6 -0.9  7
 51901.643 0.866  9.2  0.4 -0.2  7
 51902.021 0.876 14.2  1.2  4.5  2
 51902.021 0.876 11.3  1.2  1.6  2
 51902.022 0.876 13.8  1.2  4.1  2
 51919.488 0.311  1.4  2.4  0.0  7
 51919.506 0.312  2.6  1.4  1.2  7
 51954.413 0.182  7.7  1.6  0.3  7
 51954.423 0.182  4.0  2.8 -3.4  7
 51954.512 0.184  7.5  1.9  0.2  7
 51954.524 0.185  5.9  1.3 -1.4  7
 51954.538 0.185  6.0  1.6 -1.3  7
 51968.455 0.532 -2.5  2.2  0.8  7
 51968.602 0.536 -2.2  1.7  1.0  7
 52000.362 0.327 -2.2  0.9 -2.9  7
 52000.377 0.328 -3.3  0.9 -4.0  7
 52001.352 0.352 -3.8  0.7 -3.5  7
 52001.384 0.353 -1.9  0.7 -1.5  7
 52009.403 0.553 -4.6  0.9 -1.6  7
 52018.361 0.776  8.0  2.8  2.5  7
 52018.370 0.776  8.8  4.8  3.3  7
 52030.361 0.075  8.9  1.0 -2.2  7
 52031.310 0.099 14.1  1.5  3.6  7
 52031.317 0.099 16.8  2.4  6.3  7
 52032.345 0.125 13.3  2.3  3.6  7
 52032.353 0.125 13.4  1.9  3.7  7
 52339.495 0.782  6.9  0.6  1.1  7
 52363.372 0.377  0.7  0.5  1.9  7
 52363.377 0.377  0.0  2.1  1.2  7
 52364.384 0.402  1.2  0.2  3.2  7
 52367.374 0.477 -1.7  0.4  1.6  7
 52368.382 0.502  0.3  2.9  3.7  7
 52369.352 0.526 -1.1  1.5  2.2  7
 52370.437 0.553 -7.6  1.1 -4.6  7
 52370.447 0.554  1.4  5.7  4.4  7
 52386.300 0.949 17.8  2.6  6.3  7
 52389.301 0.024 12.2  0.7  0.4  7
 52396.317 0.198  6.5  2.8 -0.2  7
 52397.310 0.223  4.2  2.5 -1.3  7
 52398.338 0.249  2.1  2.8 -2.2  7
 52402.334 0.348 -2.9  3.2 -2.7  7
 52404.309 0.398 -2.6  2.6 -0.7  7
 52406.312 0.448 -1.5  2.0  1.5  7
 52407.308 0.472 -2.6  3.4  0.7  7
 52408.314 0.498 -1.6  1.9  1.8  7
 52411.341 0.573  0.8  1.6  3.4  7
 52638.871 0.245  7.1  0.8  2.6 12
 53004.861 0.369  1.8  1.0  2.8 13
 53039.774 0.240  7.7  0.9  3.0 13
 53075.679 0.135  8.1  0.9 -1.2 10
 53084.676 0.359 -0.3  0.9  0.3 10
 53084.796 0.362 -2.3  0.9 -1.6 10
 53103.639 0.832  6.2  0.9 -1.8 10
 53107.682 0.933  9.7  0.9 -1.6 10
 53111.641 0.032  9.1  1.0 -2.7 10
 53117.685 0.182  5.0  1.0 -2.4 10
 53119.683 0.232  3.5  1.2 -1.6 10
 53123.648 0.331  4.7  1.1  4.2 10
 53125.581 0.379 -0.9  1.3  0.4 10
 53129.595 0.479 -4.7  1.1 -1.4 10
 53138.614 0.704  0.5  1.1 -1.6 10
 53145.582 0.878  5.3  1.1 -4.5 10
 53152.620 0.053  9.1  0.9 -2.4 10
 53168.591 0.451 -8.1  1.4 -5.0 10
 53292.008 0.528 -9.3  2.0 -6.0  3
 53292.013 0.528 -8.2  2.0 -4.9  3
 53292.016 0.528 -5.2  2.0 -1.9  3
 53292.017 0.528 -8.4  2.0 -5.1  3
 53292.019 0.528 -8.2  2.0 -4.9  3
 53292.021 0.528 -8.7  2.0 -5.4  3
 53293.016 0.553 -6.7  1.9 -3.7  3
 53293.018 0.553 -6.5  2.1 -3.5  3
 53293.019 0.553 -6.9  2.0 -3.9  3
 53293.021 0.553 -6.7  2.0 -3.7  3
 53295.019 0.603 -2.4  2.6 -0.5  3
 53295.021 0.603 -3.3  2.6 -1.4  3
 53295.022 0.603 -5.4  2.0 -3.5  3
 53295.024 0.603 -5.1  2.7 -3.2  3
 53295.025 0.603 -4.8  2.2 -2.9  3
 53295.026 0.603 -8.5  2.1 -6.6  3
 53405.772 0.364 -0.6  1.1  0.2 10
 53426.701 0.886  7.9  1.2 -2.1 10
 53428.789 0.938 12.7  1.3  1.3 10
 53445.666 0.359  2.6  1.2  3.2 10
 53458.653 0.683 -2.7  1.2 -3.8 10
 53460.651 0.732  1.8  1.1 -1.6 10
 53465.694 0.858  9.2  1.1  0.1 10
 53466.718 0.884 11.4  1.3  1.4 10
 53496.636 0.629 -1.9  1.1 -0.9 10
 53528.577 0.426 -1.4  1.2  1.2 10
 53529.596 0.451 -4.9  1.4 -1.8 10
 53685.005 0.325  1.2  1.1  0.4  4
 53689.034 0.426 -1.9  1.1  0.7  4
 53772.781 0.514 -1.7  0.8  1.7 14
 53816.668 0.608  3.7  1.2  5.4 10
 53845.674 0.331  3.6  1.3  3.1 10
 53859.639 0.679  1.6  1.2  0.6 10
 53860.645 0.704  0.4  1.4 -1.7 10
 53878.565 0.151  6.5  1.3 -2.2 10
 53882.610 0.252 -1.2  1.5 -5.4 10
 53883.617 0.277  0.8  1.9 -2.2 10
 53892.575 0.500 -0.2  1.4  3.2 10
 53904.589 0.800  4.4  1.8 -2.2 10
 53914.592 0.049  9.6  2.1 -2.0 10
 54020.987 0.702 -3.3  2.7 -5.3  5
 54020.988 0.702 -0.9  1.9 -2.9  5
 54032.015 0.976 12.7  1.2  0.9  6
 54032.016 0.976 12.3  1.1  0.5  6
 54137.731 0.612  1.4  0.5  3.0 14
 54154.761 0.036 12.9  0.5  1.2 14
 54155.696 0.060  5.1  1.6 -6.3 10
\end{verbatim}
}

\end{document}